\documentclass[12pt,reqno,a4paper]{article}
\usepackage[T1]{fontenc}

\usepackage{authblk,amsthm,amsmath, amsfonts, caption, multirow, float, graphicx, tikz, pdfpages, setspace, enumitem, fancyhdr, longtable, accents} 
\usepackage[symbol]{footmisc}
\usepackage[misc]{ifsym}
\usepackage[normalem]{ulem}
\usepackage[round]{natbib}
\usetikzlibrary{arrows,shapes,shapes.geometric}
\usepackage{pdflscape} 
\usepackage[colorlinks,citecolor=blue,urlcolor=blue]{hyperref}
\usepackage[margin=2.5cm,headsep=10mm,footskip=15mm]{geometry}
\usepackage{pdfpages}
\usepackage{float}
\usepackage{multirow}
\usepackage{ragged2e}
\usepackage{caption}
\usepackage{subcaption}
\usepackage{hyperref}
\usepackage{xcolor}
\definecolor{green1}{rgb}{0.0, 0.5, 0.0}
\definecolor{chocolate}{rgb}{0.48, 0.25, 0.0}
\definecolor{dukeblue}{rgb}{0.0, 0.0, 0.61}
\colorlet{brown1}{brown!70!black}
\colorlet{blue1}{blue!70!black}
\colorlet{notgreen}{blue!50!yellow}
\usepackage{array}
\newcolumntype{L}[1]{>{\raggedright\let\newline\\\arraybackslash\hspace{0pt}}p{#1}}
\newcolumntype{X}[1]{>{\raggedright\let\newline\\\arraybackslash\hspace{0pt}}m{#1}}
\newcolumntype{C}[1]{>{\centering\let\newline\\\arraybackslash\hspace{0pt}}m{#1}}
\newcolumntype{R}[1]{>{\raggedleft\let\newline\\\arraybackslash\hspace{0pt}}m{#1}}

\usepackage{soul}

\sethlcolor{pink} 
\usepackage[framemethod=tikz]{mdframed}

\mdfdefinestyle{myFigureBoxStyle}{backgroundcolor=black!5,, roundcorner=3pt,tikzsetting={draw=black, line width=1pt}}%

\makeatletter
\newcommand\fs@myRoundBox{\def\@fs@cfont{\bfseries}\let\@fs@capt\floatc@plain
	\def\@fs@pre{\begin{mdframed}[style=myFigureBoxStyle]}%
		\def\@fs@mid{\vspace{\abovecaptionskip}}%
		\def\@fs@post{\end{mdframed}}\let\@fs@iftopcapt\iffalse}
\makeatother

\theoremstyle{plain}

\newtheorem*{result*}{Result}

\theoremstyle{definition}

\newtheorem*{example*}{Example}

\newcommand{\eps}{\epsilon}

\newcommand{\z}{\ensuremath{\textbf{\emph{z}}}}

\newcommand{\tbf}[1]{\noindent\textbf{#1}}

\title{Size biased Multinomial Modelling of detection data in Software testing} 

\author[1]{Pallabi Ghosh\footnote{\tbf{E-mail}: \href{mailto:pallabi26.12@gmail.com}{\textit{pallabi26.12@gmail.com}}}}
\author[2]{Ashis Kr. Chakraborty\footnote{\tbf{E-mail}: \href{mailto:akchakraborty.isi@gmail.com}{\textit{akchakraborty.isi@gmail.com}} }}
\author[3]{Soumen Dey\footnote{\tbf{E-mail}: \href{mailto:soumenstat89@gmail.com}{\textit{soumenstat89@gmail.com}} }}

\affil[1]{Department of Statistics, St. Xavier's College (Autonomous), Kolkata}
\affil[2]{SQC \& OR Unit, Indian Statistical Institute, Kolkata}
\affil[3]{Norwegian University of Life Sciences, \r{A}s , Norway}

\begin{document}
\tikzstyle{every picture}+=[remember picture] 
		
\everymath{\displaystyle}

\setlength{\abovedisplayskip}{5pt}
\setlength{\belowdisplayskip}{5pt}
\date{}
\maketitle

\tbf{Running headline}: Estimation of Software Reliability


\vspace{10cm}
\pagebreak

	


{\footnotesize
\begin{center}	{\bf Abstract } \end{center}
Estimation of software reliability often poses a considerable challenge, particularly for critical softwares. Several methods of estimation of reliability of software are already available in the literature. But, so far almost nobody used the concept of size of a bug for estimating software reliability. In this article we make used of the bug size or the eventual bug size which helps us to determine reliability of software more precisely. The size- biased model developed here can also be used for similar fields like hydrocarbon exploration. The model has been validated through simulation and subsequently used for a critical space application software testing data. The estimated results match the actual observations to a large extent. 


\vspace{1cm}

\tbf{Keywords}: size biased, software reliability, Bayesian method, critical software, eventual size of a bug.
} 




\section{Introduction}\label{sec:intro}

\par Estimation of software reliability has been a very important and interesting problem being tackled for almost half a century now. Still newer challenges are offered to the researchers because of the variety of software testing data in different environments. Commercial software, in many cases, are not stand-alone software. Hence, the performance in terms of reliability of the software cannot be judged as it had been done earlier. Similarly, software used in missions are also not stand-alone software. Moreover, due to substantial effect of a failure in space, software used in space research needs to be highly reliable.\\

A variation of the software reliability problem is to find out optimum testing time of software based on certain reasonable criteria.Several researchers addressed these problems in different ways and under different assumptions. Yamada and Osaki ( 1985)\cite{yamada1985osaki}, Chakraborty and Arthanari (1994)\cite{Chakraborty1994Optimum}, Nayak (1988) \cite{Nayak1988Estimation} and a few others proposed to use software testing data as it appears in a discrete set up. They suggested optimal testing time and reliability estimation of a software under discrete framework.

For large software, testing data are generally logged in test-case wise, irrespective of whether it is individual module testing or integrated testing for the whole softwre (Dewanji et al. (2011)) \cite{Dewanji2011Reliability}.

There are a few literature on this discrete set up for software reliability estimation. Notable among them are Yamada and Osaki ( 1985) \cite{yamada1985osaki}, Chakraborty and Arthanari (1994)\cite{Chakraborty1994Optimum}, Chakraborty (1996)\cite{Chakraborty1996Software}, Chakraborty et. al (2019)\cite{Chakraborty2019Bayesian}, Dewanji et. al (2011) \cite{Dewanji2011Reliability} and a few others. Initial developments in discrete set up was mostly based on some assumptions which were found to be not applicable for most software testing data (Chakraborty (2019))\cite{Chakraborty2019Bayesian}. Baker (1997) \cite{Baker1997} developed two growth models by considering suitable probability distributions for the number of faults initially present per module. The author used successive testing of software modules till no further faults were found. However, it is well known that it is not safe to announce that a software or a module does not have any error in it, even if it is tested many times.

Worwa (2005) \cite{worwa2005} proposed a model to determine the mean value of the number of errors encountered
during the program testing process assuming the program (software/ module) under testing is characterized by characteristic matrix (($p_{mn}$)), where $p_{mn}$ is the probability of detecting `$n$'  errors encountered during a single stage (phase) of the program testing if there are `$m$' tests (test runs) that have incorrect execution
in that stage(Dewanji et al.(2011))\cite{Dewanji2011Reliability}. Particularly for space research the authors suggested some software reliability growth model under discrete framework, which they have called as ‘periodic debugging’. They used logistic regression to model the probability of a failure (run with error) or success (error-free run) and the method of maximum likelihood estimation to estimate the model parameters. In this work, they also suggested a method to determine the minimum number of error-free test runs required additionally for estimated reliability to achieve a specific target reliability with some high probability. Chakraborty et al. (2019)\cite{Chakraborty2019Bayesian} also generalised their earlier model (1994) \cite{Chakraborty1994Optimum} for optimum stopping rule at release time under discrete periodic debugging setup. In this paper they proposed a reward function for observing and debugging an error and derived, using Bayesian approach, an optimum stopping rule of software testing using a function called `maximum expected return’ based on ‘reward’ and cost incurred for each run of a test case.

\par JIA et al.(2010)\cite{jia2010} used markov chain to estimate software reliability if the software is tested additional $T$ units of time. Wang et al.(2016) \cite{wang2015} developed a model to analyze the time dependencies between the fault detection and correction processes. They also proposed another NHPP model using number of
fault detection corrected by time `t' as well as under correction at time `t'. In this work, two separate likelihood functions were considered for fault detection process (FDP) and fault correction process (FCP) respectively. Again a general form of a joint likelihood function was also developed for combined FDP and FCP to estimate the software fault detection intensity and fault correction intensity. The model parameters were estimated by maximum likelihood method of estimation. Li and Pham (2021)\cite{Li2022} have also focused on both the processes considering
the situation of imperfect debugging based on NHPP. The authors used three different testing coverage (code percentage that has been examined up to time t) functions to develop three separate models considering the dependency
between fault detection and correction processes by a ratio of the cumulative corrected faults and the cumulative detected faults up to time t. Li and Pham have used the method of least squares to estimate the model parameters and
combined mean squared error (MSE) and mean relative error (MRE) were used for model fitting.

\par Finally, a software can be considered as a tree structure having many paths and sub-paths completing the full tree. When testing a software, some bugs are detected very quickly and some bugs are very difficult to get detected. The procedure of testing a software normally is done through phases where in each phase a specific, but different number of test cases are tested during simulation testing \cite{Dewanji2011Reliability}. Much before that, each of the modules are also tested using specially developed test cases and even before that independent experts go through the software code, known as walk through, to identify syntax and other obvious errors in the software. So, over all, there are three steps in which faults or errors in software are detected and in each step, it is recorded. In fact, for a critical software, whenever any change in the code takes place, all the three above steps are conducted to ensure that no new faults are added in the software. However, since nobody can guarantee that no bugs are there in any software, estimating reliability of software is probably the only way to have confidence in the software \cite{Pham2000Reliability}. Quite clearly, considering the tree structure of the software, a bug on a common path is more likely to be detected earlier than a bug which is on a subpath or a sub-sub path, because the chances that an input will come across the earlier bug is more compared to the bug that exists on the sub-path or sub-sub path. This natural phenomenon is used in our modeling and is defined as size-biased modeling \cite{Patil1978sizebiased}. In the following subsection we will define the concept clearly before proceeding to the modelling section.

\par The procedure of testing software goes through some phases and each phase has specific test cases. If a bug is on a very common path (that is, most of the test cases are likely to pass through the bug), then it is easier to detect the bug quickly, but if the bug is on some rare path, then the test cases may fail to detect the bug. If many inputs (test cases or other inputs) are likely to pass through a bug, then the size of the bug can be assumed to be larger compared to the bugs that are traversed by a relatively fewer inputs. But if with a few test cases one can detect a bug, then the size of the bug may be assumed to be small. On the basis of this concept, we define the size of a bug as the total number of inputs, which consists of test cases as well as the inputs that may be used by future users which will go through a particular bug. Even though after debugging, the bug may not exist, but a count of such inputs which otherwise would have detected the bug forms the eventual size of a bug.

\par We will be using this new bug size concept for approximately estimating reliability of the software. The novel size biased sampling concept is probably mentioned for the first time in \cite{Chakraborty1994Optimum} followed by \cite{dey2022estimating} and it provides a very accurate estimate of reliability of the ISRO flight control software.

\par Also going by the same concept, just knowing the number of bugs remaining in the software after testing is over, is not enough unlike most of the research carried out on software reliability. The reliability of the software can be better judged or estimated, if we can estimate the total remaining size of the bugs (that is, total eventual size of the remaining bugs) that remain undetected even after testing, that is, the total estimated number of inputs that may find some bugs in the software even after testing. This logically is a better measure of software reliability than a measure which is based on the remaining number of bugs. It may be noted that the remaining total size of the bugs is always greater than or equal to the remaining number of bugs. Since the number of inputs that may encounter such bugs after testing is completed is not known any time, we need to estimate this and we call this as the total eventual size of the remaining bugs. If this total eventual remaining bug size is small, the software is more reliable, otherwise, the software remains unreliable.

\par Taking cue from this discussion, we define software reliability as the probability that the total eventual remaining size is less than a given quantity, say $\epsilon$. The value of $\epsilon$, even though is fixed arbitrarily, may be obtained after discussing with the concerned software engineers. Also, after several successful missions, one may reduce the value of $\epsilon$, providing more confidence on the software.

\par Some important features of the ISRO (Indian Space Research Organization) flight control software is that the data collected are discrete in nature , in terms of whether a test case could identify a bug or not. More about ISRO flight control software data is given in Section 7. Other important features include 1) different test cases are used in different phases of simulation testing and we have considered eight such phases. Also, 2) debugging of all bugs identified in a phase are done at the end of the corresponding phase. 3) Each phase generally is applied to test different functionals in a software.

\par These important features of ISRO flight control software can be found in other critical and large softwares as well. Hence, the model developed in this article is applicable elsewhere as well. The data collected before each mission covers basically the number of inputs (or test cases as they are called) used for each phase of testing and how many bugs used for each phase of testing and how many bugs could be identified for each such testing. Similar data are available for each module testing also. However, mistakes found during walk through are also documented but not used for modeling purposes.\vspace{2mm}
\\
In this article we try\vspace{1mm}
\\
1) to estimate total number of bugs present in the software.\vspace{2mm}
\\
2) to estimate the remaining eventual bug size, that is, to estimate the total eventual size of the remaining bugs. \vspace{2mm}
\\
3) to estimate the reliability of software at release based on the above.

\section{Modeling}\label{sec:modeling}


Let there be $J$ number of missions (known), $K$ number of phases (known) and $N$ number of bugs (unknown) in a software used specifically for space programs.

Let us assume that the maximum number of possible bugs be $M$. We introduce a variable $z_{i}$ that takes the value 1 if $i^{th}$ bug is real and takes the value 0 otherwise, $i=1,2,\dots,M$. So sum of these binary variables gives the total number of bugs, that is,  $N = \sum_{i=1}^M z_i$. Further, we assume $z_i \sim \text{Bernoulli}(\psi)$, where $\psi$ is the probability that a bug is real. Let $S_{i}$ be the eventual size of $i^{th}$ bug and $\alpha_{i}$ be the probability that $i^{th}$ bug is detected on any one of the mission at any of the phases. It is clear from the above discussion that a bug with larger size should have higher detection probability. Therefore, the detection probability $\alpha_{i}$ should be modelled as an increasing function of the bug size $S_{i}$.
\par
If $i^{th}$ bug is real, i.e., $z_i=1$, consider another binary variable $w_{ijk}$ which takes the value 1 if $i^{th}$ bug is detected in the $j$-th mission at $k$-th phase and takes the value 0 otherwise, $j=1,2,\dots,J$, $k=1,2,\dots,K$. Also note that, for the ISRO data set, when $ z_i = 1 $, we have $\sum_{j=1}^J \sum_{k=1}^K w_{ijk} \leq 1$,  for each $i = 1,2,\dots,M$, since some of the bugs may not get detected and we will have $w_{ijk}=0$ for each mission $j$ and phase $k$. Further, we introduce a binary variable $w_i^*$ which takes the value 1 when the $i$-th bug is not captured at any of missions and phases and takes 0 otherwise.
\par
The variable $w_{ijk}$ is actually a categorical variable. That means, for a real bug with $z_i = 1$, we have the distribution of $w_{i11}, w_{i12}, \dots, w_{iJK}, w_i^*$ conditional on  $S_{i}$ as
\begin{align}
(w_{i11}, w_{i12}, \dots, w_{iJK}, w_i^*) \,| \, S_{i} \sim \text{Multinomial}(1, \, \{\alpha_i \, p_{11}, \alpha_i \,p_{12}, \dots, \alpha_i \,p_{JK}, 1-\alpha_i\}),
\end{align}The above conditional distributions are independently distributed for each $i$ = $1,2,\dots,M$. If the bug is not real, we have $z_i=0$ and consequently $w_{ijk} = 0$ with probability 1 for each $j = 1,2,\dots,J$ and $k=1,2,\dots,K$.  Here, $p_{jk}$ be the probability of detecting a bug in $j^{th}$ mission and $k^{th}$ phase. Let $T_{jk}$ be the no. of test cases used in $j^{th}$ mission at $k^{th}$ phase. Note that, probability of detecting a bug in a particular phase of a mission should increase as no. of test cases used  in that phase of the mission increases. Therefore, we can express $p_{jk}$ as: $p_{jk} = 1 - \exp \, (-T_{jk})$. Again, we can write $\alpha_{i} = 1 - \text{Pr}\, (\text{$i^{th}$ bug is not detected in any of these missions at any of the phases})$. Now, as we have discussed before, that a bug with larger size has greater chance of being detected, $\alpha_{i}$ (i.e., the probability of $i^{th}$ bug being detected in any of the missions at any of the phases) is also proportional to $S_{i}$. Similarly we can say that a bug with larger size will have lesser probability of non-detection. Therefore, we can consider, 
\begin{align}\label{detkernel}
\text{Pr}(\text{$i^{th}$ \small{bug is not detected in any missions at any of the phases}}) = \exp(-\frac{S_{i}^\nu}{\underset{j,k}{\max}\{T_{jk}\}})
\end{align} ,
where $\nu$ is a tuning parameter which controls the rate of decay in the above detection probability kernel (\ref{detkernel}). 
Consequently, $\alpha_i$ can be expressed as: $$\alpha_i = 1 -\exp \, (-\frac{S_{i}^\nu}{\underset{j,k}{\max}\{T_{jk}\}})$$. 
\par
So, from the above multinomial model, we are implicitly assuming that the probability of detecting a bug with a single test case varies over different missions and different phases.
\par
Also the count $y_{jk}= \sum_{i=1}^M w_{ijk} \, z_i$ gives the total number of bugs detected in the $j$-th mission at $k$-th phase. The count data $y_{jk}$ can be viewed as reduced information summaries of the parental (latent) data $w_{ijk}$ that would be observed if all bugs in the software were marked or distinguishable in the given data set. Then the distribution of $(y_{11}, y_{12}, \dots, y_{JK}, w_0^*)$ conditional on $z_i$ is
\begin{align}
(y_{11}, y_{12}, \dots, y_{JK}, w_0^*) \sim \text{Multinomial}(N, \, \{\alpha_i \, p_{11}, \alpha_i \,p_{12}, \dots, \alpha_i \,p_{JK}, 1-\alpha_i\}).
\end{align} 
The variable $w_0^* = \sum_{i=1}^M w_i^* \, z_i $ gives the count of number of bugs not detected out of the $M$ possible bugs.
 Note that, there are only $N$ terms involving $w_{ijk}$ in the sum expression: $ \sum_{i=1}^M w_{ijk} \, z_i $ since the other $M-N$ term vanishes due to $z_i=0$ for those non-existing bugs.

\subsection{Methods}\label{sec:methods}
We use Bayesian method to estimate the parameters. We first set a value for $M$. Let the number of detected bugs be $n$. This is clear that $w^{\ast}_{i}$ takes 0 for the $n$ detected bugs and takes the value 1 for $(M-n)$ bugs. Clearly, these $n$ bugs are real. However, $(N-n)$ bugs are also real, but not yet identified. So, $z$ takes the value 1 for the $n$ detected bugs and also for the $(N-n)$ undetected but, real bugs. But we can not assign the values of $z$ for the remaining $(M-n)$ bugs as the total number of bugs ($N$) is unknown. So, we generate random observations from the posterior distribution of $\z_{i}$ assuming some initial choices of $\psi$ and estimate $N$ using $N=\displaystyle\sum_{i=1}^{M} z_{i}$. Once we get an estimate of $N$, we can draw sample from posterior distribution of $\psi$, and hence from the posterior distribution of $S_{i}$ and $\lambda_{i}$. Thus we can estimate the detection probability $\alpha_{i}$, as $\alpha_{i}$ is a function of $S_{i}$. On the basis of the updated estimates of $\psi$ and $\alpha_{i}$ we repeat the procedure until a sufficient number of chains are generated. We use posterior mean to estimate $N$, $\psi$, $\{S_{i}: i = 1,2, \dots,M\}$ and $\{\lambda_{i}$ : $i=1,2,\dots,M\}$. 

\subsection{Estimating Reliability}\label{sec:reliability}
If $S_{i}$ is the eventual size of $i^{th}$ bug and $z_{i}$ indicates if the $i^{th}$ bug is real or not, then $\displaystyle\sum_{i=1}^{M} S_{i}z_{i}$ explains the total eventual size of all the bugs in the software. Since, $u_{i}$ has been declared as 1 if the $i^{th}$ bug is not detected, and 0 otherwise, $\displaystyle\sum_{i=1}^{M} S_{i}(1-u_{i})$ is the eventual size of detected bugs. Therefore, the eventual size of the remaining bugs can be estimated by $R = \displaystyle\sum_{i=1}^{M} S_{i}z_{i} - \displaystyle\sum_{i=1}^{M} S_{i}(1-u_{i})$.\vspace{2mm}
\\
Note that, the remaining size should be small enough as it is desirable that the bugs with larger size are already detected. So, based on this concept the reliability of the software can be defined by $\text{Pr} \, (R \, < \, \eps)$, for a given value of $\eps$. With the help of simulation we can estimate the reliability of the software.



\subsection{Prior specification}\label{sec:prior}

Eventual bug sizes ($S_{i}$'s) are usually latent and unobservable. We assign a negative binomial-gamma mixture prior for $S_{i}$ to capture the required level of variability in the latent variable. Consequently, each $S_i$ is assumed to follow negative binomial distribution with mean $\lambda_i$, where the $\lambda_i$ is a random draw from gamma distribution with shape parameter $a_s$ and rate $b_s$ and to make the prior non-informative we choose the parameters such that the variance of $\lambda$ is high:
\begin{align}
   &  S_{i} \, | \, \lambda_{i} \sim \text{Negative binomial}(mean=\lambda_{i}), \nonumber\\
   &  \lambda_{i} \sim \text{Gamma}(a_s=50,\,b_s=0.5 ), 
\end{align}
for each $i = 1,2, \dots, M$.
We assign bounded Uniform prior over the interval $(0,1)$ for the inclusion probability $\psi$. That is, $\psi \sim \text{Uniform}(0, \, 1) $.

\subsection{Posterior distributions}\label{sec:posterior}

The posterior density of $[S_{i}| w_{i11},w_{i12},...w_{iJK},\lambda_{i}]$,
\begin{align}
    [S_{i}| w_{i11},w_{i12},...w_{iJK},\lambda_{i}] = \dfrac{[w_{i11},w_{i12},...w_{iJK}|S_{i}][S_{i}|\lambda_{i}]}{\sum_{S_{i}}[w_{i11},w_{i12},...w_{iJK}|S_{i}][S_{i}|\lambda_{i}]}
\end{align}
The posterior density of $[\lambda_{i}|S_{i}, w_{i11},w_{i12},...w_{iJK}]$,
\begin{align}
    [\lambda_{i}|S_{i}, w_{i11},w_{i12},...w_{iJK}] = \dfrac{[w_{i11},w_{i12},...w_{iJK}|S_{i}][S_{i}|\lambda_{i}][\lambda_{i}]}{\int_{\lambda_{i}} [w_{i11},w_{i12},...w_{iJK}|S_{i}][S_{i}|\lambda_{i}][\lambda_{i}] d\lambda_{i}} 
\end{align}
The posterior density of $[\psi|z_{1}, z_{2}, \dots, z_{M}]$,
\begin{align}
    [\psi|z_{1}, z_{2}, \dots, z_{M}] = \dfrac{\psi^{N}(1-\psi)^{M-N}}{Beta(N+1, M-N+1)})
\end{align}

\section{Simulation study}\label{sec:simstudy}
\subsection{Model fitting}
The model is fitted using Markov chain Monte Carlo (MCMC) Simulations. A data has been generated for the fixed values of the parameters. Then, on the basis of data, Metropolis - Hastings algorithm has been used to generate values of the parameters from the target posterior distributions. Here we simulate three chains each of length 50000, and then discard the first half of each. 

\subsection{Model convergence and model precision}
We use the potential scale reduction factor ($\hat{R}$) and the effective sample size to monitor the convergence of the Markov chains. The iterations should be repeated until the val $\hat{R}$ is near 1 and the effective number of independent simulation draws (effective sample size) is large enough for all quantities of interest. We also use traceplots for visual inspection of convergence.

\subsection{Simulated data}
To verify model fitting we start with a simulated data. The total number of missions and the total number of phases corresponding to each mission have been set to 30 and 8 respectively. Total no. of bugs in the software has been assumed as 100 and the no. of maximum possible bugs has been considered as 400. The data for test cases are sampled from 0 to 50. We verify the model with three different values of the tuning parameter $\nu = 1, 1.25, 1.5$.

\subsection{Results of simulation study}
Here we discuss the results for the value of $\nu$ for which the model is well-fitted. We use traceplot of three chains for visual inspection of the convergence of the model parameters $N$ (Total number of bugs in the software), $\psi$ (probability that a bug is real), $S_{i}$ (eventual size of $i^{th}$ bug) and $\lambda_{i}$ (average eventual size of $i^{th}$ bug) (see Figure~\ref{tab:my_label1}) For each chain we obtain the posterior summaries like mean, standard deviation and coefficient of variation (see Table~\ref{tab:my_label1} and Table~\ref{tab:my_label2}).

\begin{figure}[H]
\begin{subfigure}{0.55\textwidth}
   \includegraphics[page=1,width=\textwidth]{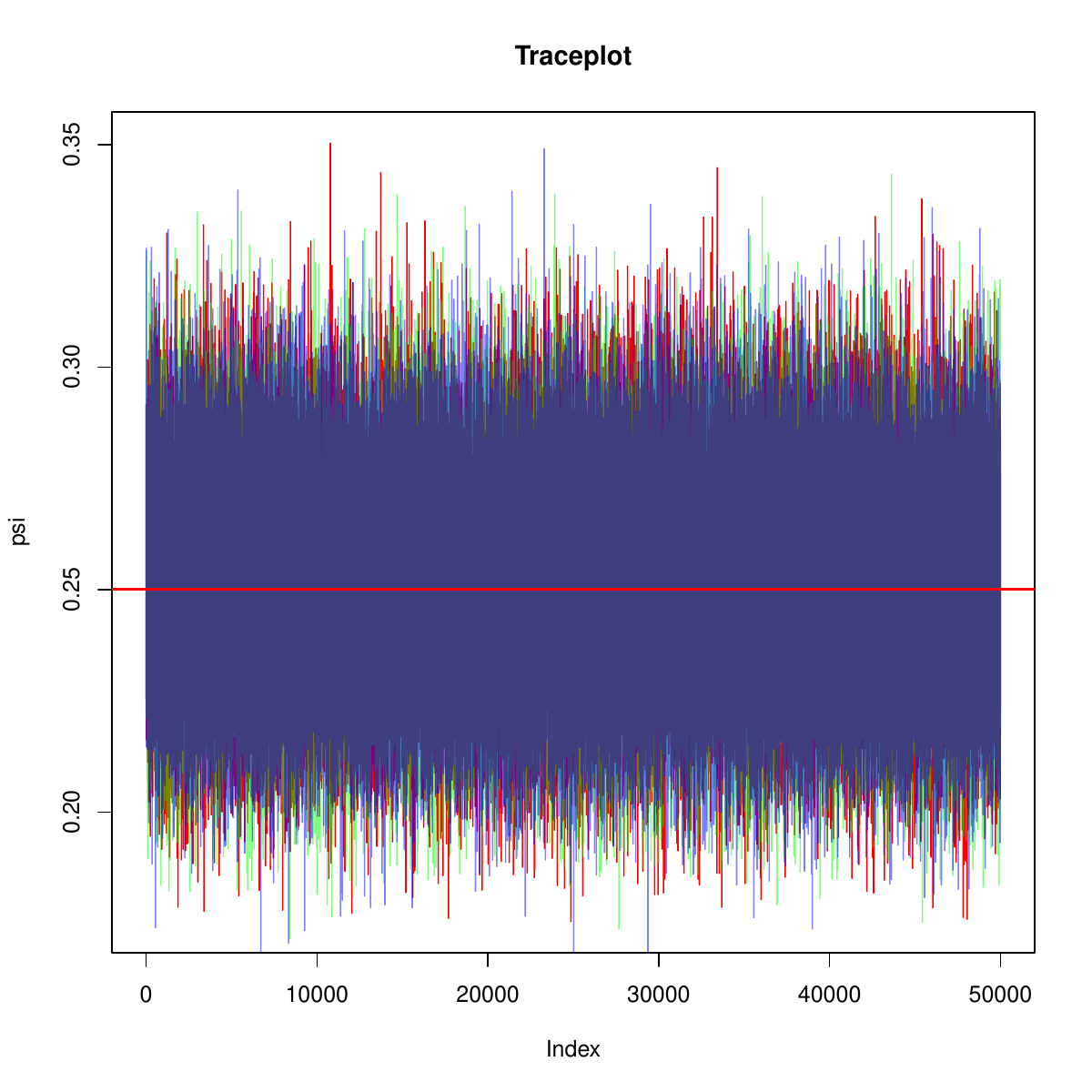}
   \caption{}
\end{subfigure}
\begin{subfigure}{0.55\textwidth}
     \includegraphics[page=2,width=\textwidth]{nu_1.5_S_NB.pdf}
     \caption{}
\end{subfigure}
\caption{(a) The traceplot $\psi$ for simulation study based on three chains each of 50000 simulations for the model with tuning parameter $\nu=1.5$ (b) The density plot of $\psi$ obtained by kernel density estimation with Gaussian kernel}
\end{figure}

\begin{figure}[H]
\begin{subfigure}{0.55\textwidth}
   \includegraphics[page=3,width=\textwidth]{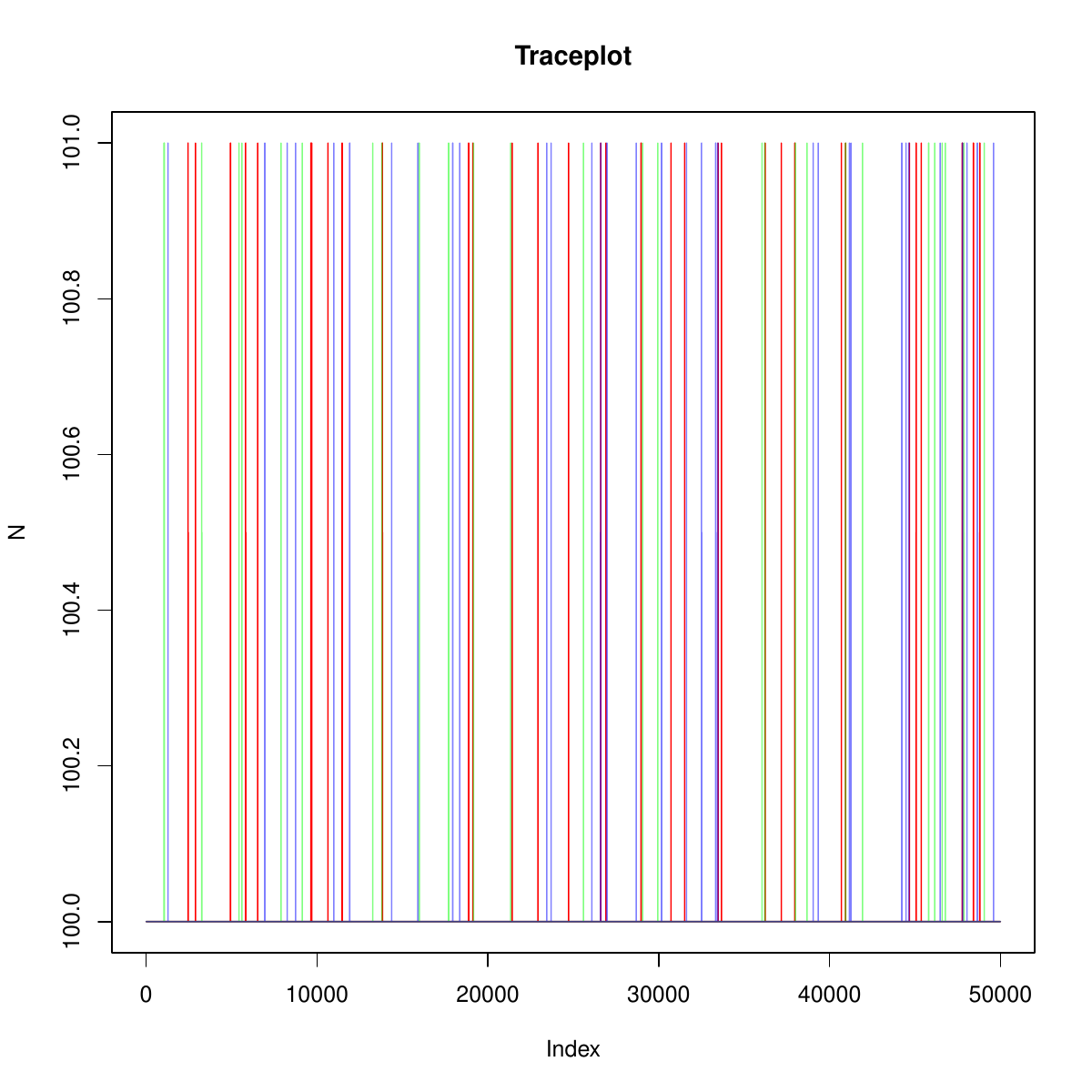}
   \caption{}
\end{subfigure}
\begin{subfigure}{0.55\textwidth}
     \includegraphics[page=4,width=\textwidth]{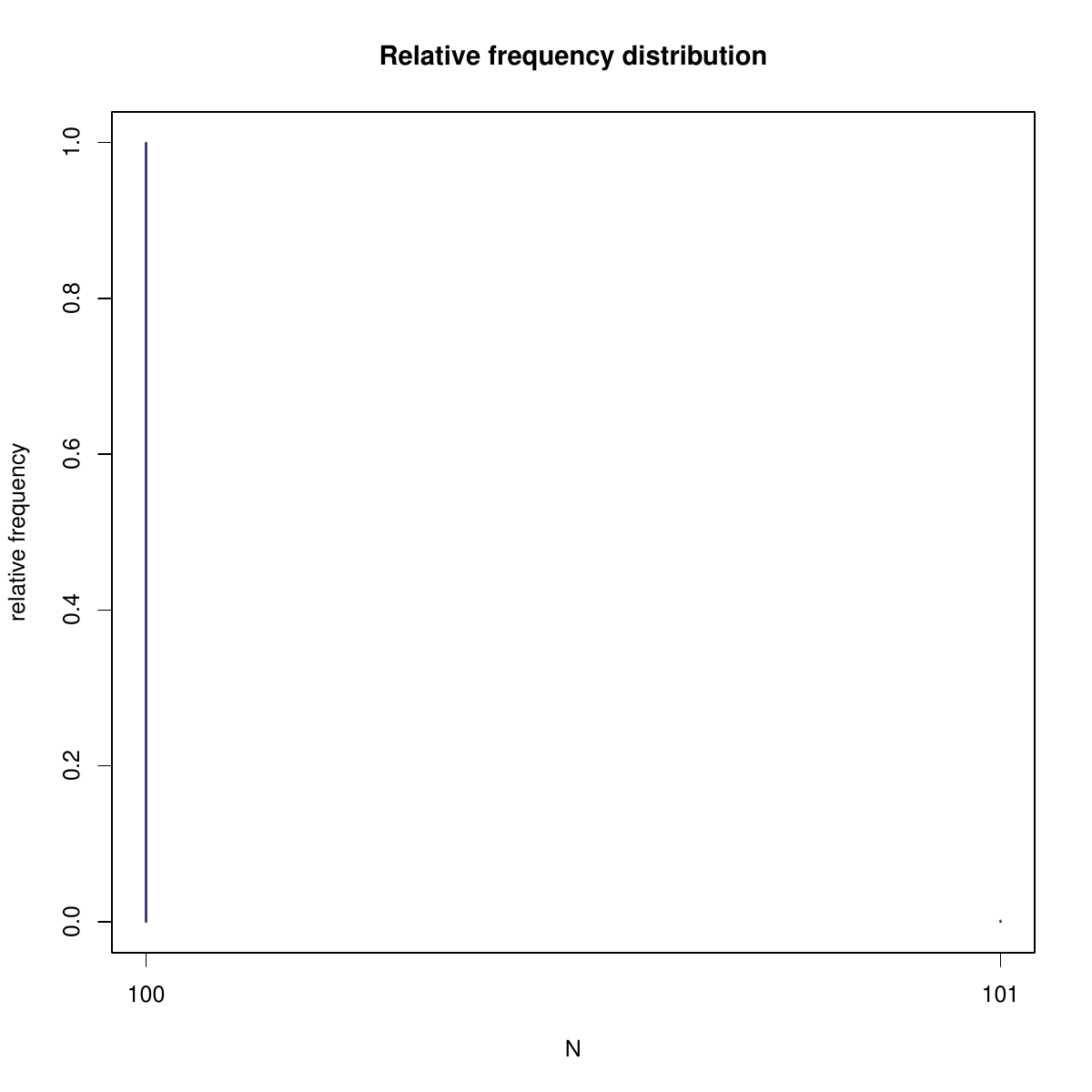}
     \caption{}
\end{subfigure}

\begin{subfigure}{0.55\textwidth}
   \includegraphics[page=15,width=\textwidth]{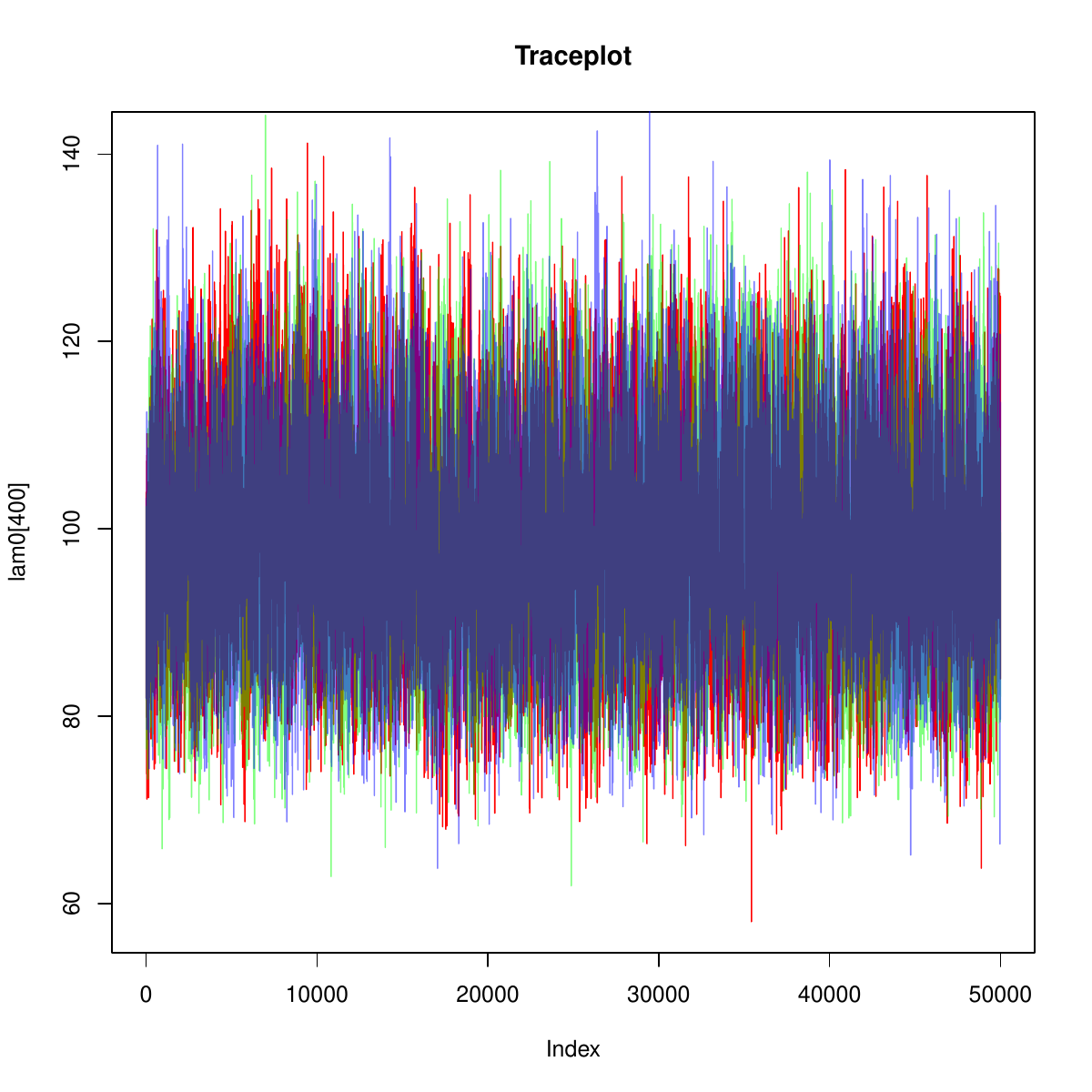}
   \caption{}
\end{subfigure}
\begin{subfigure}{0.55\textwidth}
     \includegraphics[page=16,width=\textwidth]{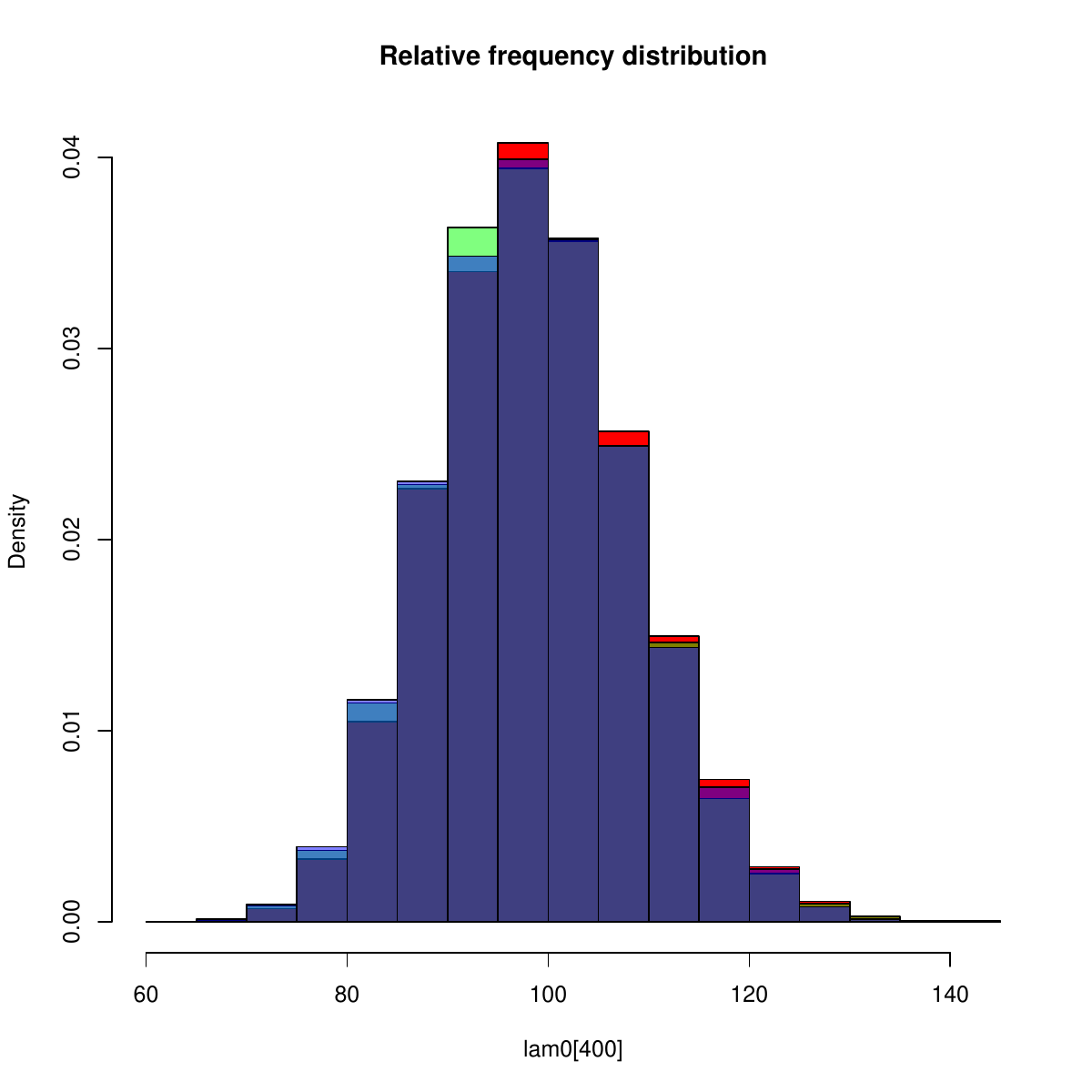}
     \caption{}
\end{subfigure}
\caption{(a) The traceplot of $N$ based on the three chains and 50000 repetitions for ISRO data (b) The sampling distribution of $N$ (c) The traceplot of $\lambda_{M}$ (d) The sampling distribution of $\lambda_{M}$. Here $M$ is set as 400}
\end{figure}

\begin{figure}[H]
\begin{subfigure}{0.55\textwidth}
   \includegraphics[page=27,width=\textwidth]{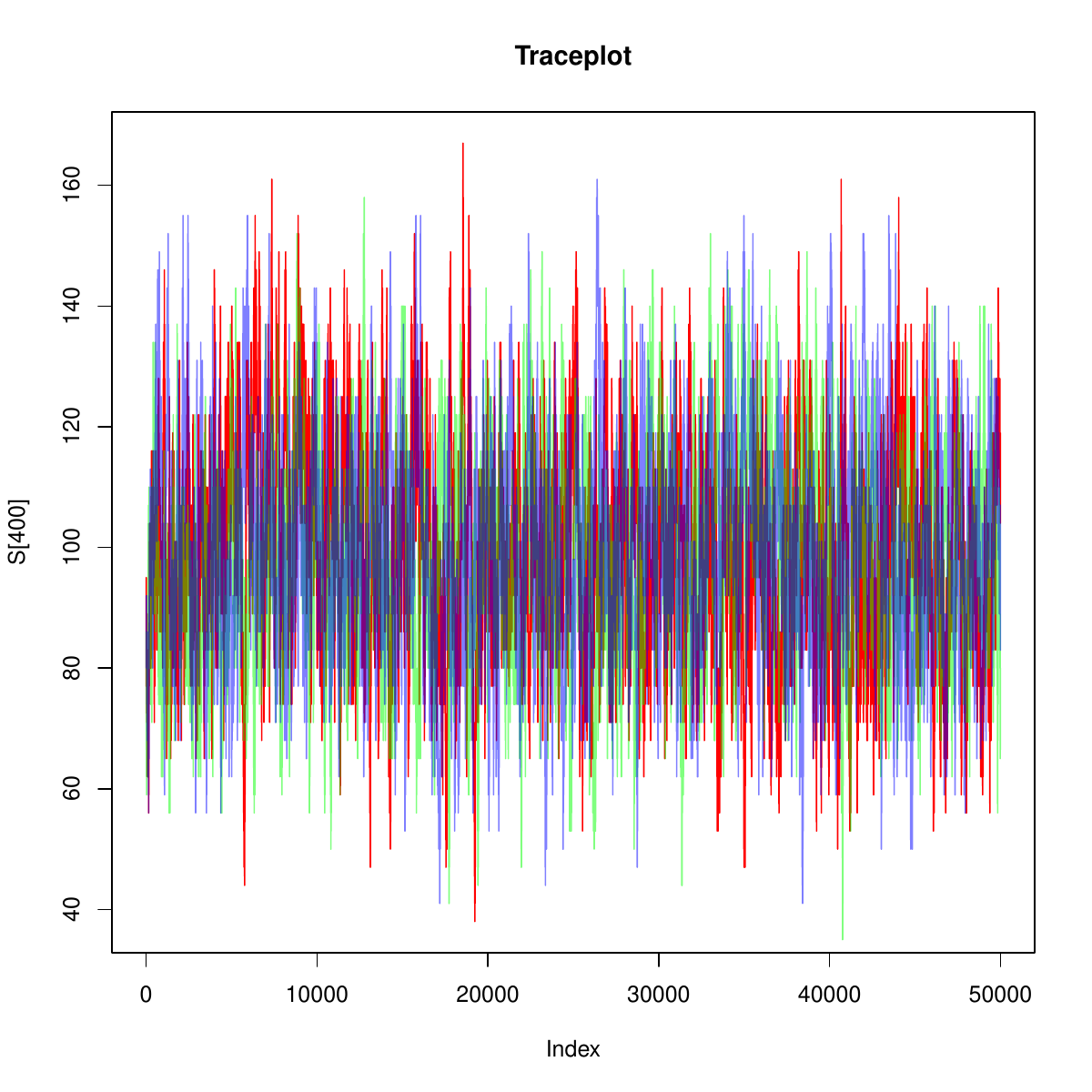}
   \caption{}
\end{subfigure}
\begin{subfigure}{0.55\textwidth}
     \includegraphics[page=28,width=\textwidth]{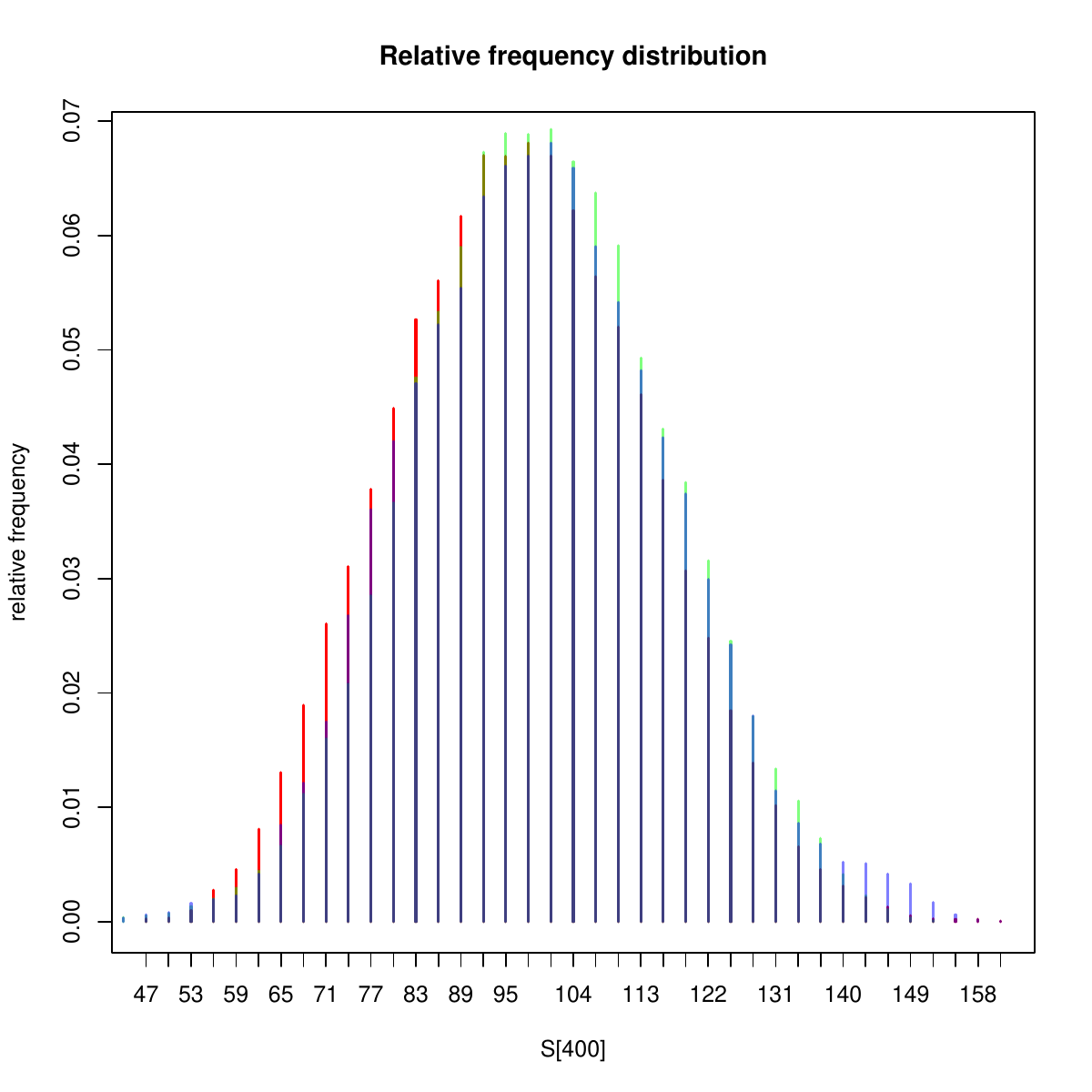}
     \caption{}
\end{subfigure}
\caption{(a) The traceplot $S_{M}$ for simulation based on three chains each of 50000 simulations (b) The sampling distribution of $S_{M}$. $M$ is set as 400.}
\end{figure}


The following table shows the descriptive measures for the model parameters:\vspace{2mm}
\\
\begin{table}[H]
    \centering
    \caption{Posterior summaries for each chain}
    \begin{tabular}{|p{1.09cm}|c|c|c|c|c|c|c|c|c|}
    \hline
        \multirow{2}{*}{para-} & \multicolumn{3}{c|}{mean} & \multicolumn{3}{c|}{sd} & \multicolumn{3}{c|}{cv}\\
         \cline{2-10}
         meters& ch1 & ch2 & ch3 & ch1 & ch2 & ch3 & ch1 & ch2 & ch3\\
         \hline
         $\psi$ & 0.2513 & 0.2511 & 0.2514 & 0.0215 & 0.0215 & 0.0214 & 8.56 & 8.58 & 8.53\\
         \hline
         $N$ & 100.0006 & 100.0007 & 100.0009 & 0.0253 & 0.0261 & 0.0303 & 0.03 & 0.03 & 0.03\\
         \hline
         $\lambda_{1}$ & 98.2524 & 99.2390 & 98.9278 & 9.7149 & 9.8999 & 9.7799 & 9.89 & 9.98 & 9.89\\
         \hline
         $\lambda_{2}$ & 99.4228 & 99.2390 & 98.9278 & 10.0447 & 9.8999 & 9.7799 & 10.10 &  9.98 & 9.89\\
         \hline
         $\lambda_{M-1}$ & 99.3942 & 99.0823 & 98.8950 & 9.6896 & 10.1509 & 9.9748 & 9.75 & 10.24 & 10.09\\
         \hline
         $\lambda_{M}$ & 98.6746 & 100.0415 & 99.0308 & 9.8630 & 10.2836 & 10.1298 & 10.00 & 10.28 & 10.23\\
         \hline
         $S_{1}$ & 96.6647 & 100.0740 & 98.4286 & 16.2287 & 17.0804 & 16.2674 & 16.79 & 17.07 & 16.53\\
         \hline
         $S_{2}$ & 100.3628 & 97.8896 & 99.0992 & 17.7097 & 16.1495 & 17.0072 & 17.65 & 16.50 & 17.16\\
         \hline
         $S_{M-1}$ & 100.0440 & 99.4108 & 98.6426 & 16.3879 & 17.8659 & 17.5607 & 16.38 & 17.97 & 17.80\\
         \hline
         $S_{M}$ & 97.8814 & 101.7416 &  98.9591 & 16.5848 & 18.0404 & 18.3715 & 16.94 & 17.73 & 18.56\\
         \hline
    \end{tabular}
    \label{tab:my_label1}
\end{table}

The estimates of the diagnostic measures to verify the convergence of the model parameters are tabulated below:\vspace{2mm}
\\
\begin{table}[H]
    \centering
    \caption{Diagnostic measures of convergence for different parameters}
    \begin{tabular}{|c|c|c|c|}
    \hline
         parameters & $\hat{R}$ & Upper C.I. & ESS\\
         \hline
         $\psi$ & 1 & 1 & 74386.52\\
         \hline
         $N$ & 1 & 1 & 63859.76\\
         \hline
         $\lambda_{M-2}$ & 1 & 1.01 & 1724.847\\
         \hline
         $\lambda_{M-1}$ & 1 & 1 & 1950.843\\
         \hline
         $\lambda_{M}$ & 1 & 1.02 & 1625.500\\
         \hline
         $S_{M-2}$ & 1.01 & 1.02 & 506.9110\\
         \hline
         $S_{M-1}$ & 1.01 & 1.02 & 525.6959\\
         \hline
         $S_{M}$ & 1.01 & 1.04 & 500.4259\\
         \hline
         
    \end{tabular}
    \label{tab:my_label2}
\end{table}

The above traceplots and the measures are given for the model with tuning parameter $\nu=1.5$. We see from the traceplots all three chains converge for all the model parameters. The estimates of the parameters are close enough with low value of coefficient of variation. The value of potential scale reduction factor ($\hat{R}$) also indicates convergence for all the parameters.

\section{ISRO Data}\label{sec:isrodata}

Here we are going to apply a flight control software data to our model. The software testing has been conducted in two main stages where test cases are used - module testing and simulation testing. Each software has finite number of modules. In module testing, each module of a software is tested on the basis of some test cases and outputs (errors) are observed. Finally integrated tests (simulation testing) are conducted in seven different phases to verify the overall performance of a software and the outputs (errors) are observed. Like module testing, in this stage (simulation testing) also some pre-specified inputs are considered for which the outputs are verified. We have used the data on module testing and simulation testing for modelling purpose. The number of inputs (test-cases) and the number of outputs (errors) have been reported for 35 missions.\vspace{2mm}
\\
\begin{table}[H]
    \centering
    \caption{Sample data of test cases from ISRO software testing}
    \begin{tabular}{|c|p{1.5cm}|p{1.5cm}|p{1.5cm}|p{1.5cm}|p{1.5cm}|p{1.5cm}|p{1.5cm}|}
    \hline
    \multirow{2}{*}{Missions} & \multicolumn{2}{c|}{Phase 1} & \multicolumn{2}{c|}{Phase 2} & \dots & \multicolumn{2}{c|}{Phase 8}\\
    \cline{2-8}
    & number of bugs & number of test cases & number of bugs & number of test cases & \dots & number of bugs & number of test cases\\
     \hline
     M1 & 3 & 61	& 0 & 10 & \dots & 0 & 38\\
     \hline
     M2 & 9 & 59 & 0 & 10 & \dots & 0 & 65\\
     \hline
     \vdots & \vdots & \vdots & \vdots & \vdots & \vdots & \vdots & \vdots\\
     \hline
    \end{tabular}
    \label{tab:my_label3}
\end{table}

\subsection{Application to ISRO Data}\label{sec:empiricalstudy}

We select the tuning parameter as $\nu = 1.5$ in the previous section by simulation study. Now, we apply the to the ISRO software testing data. 
\par
We frame the data of software testing so that it fits to our model. As in the module testing, each module of a software is tested, we count the total number of bugs observed in all the modules of a software and assume it as the first phase. After module testing, simulation testing is performed in seven different phases. So, there are altogether eight phases to test a software for a particular mission. The observations are reported for 35 missions. Total 61 bugs were observed in the data. To fit data to our multinomial model, an array of dimension $M \times (JK+1)$ has been considered, where each row indicates observations (that is, if a bug is detected) corresponding to a bug and the value is 1 only when the bug has been detected in any phase of a mission. That is, if the first bug has been detected in the first mission at phase 1, the value will be 1 only for that phase of the mission and the value will be 0 for all other phases of the mission. For an undetected but real bug, all cells will take 0. For non-existent bugs as well all cell values will be zero. So, the sum of any row is either 0 or 1. 
The results for different tuning parameters are attached as appendix.

\subsection{Results}\label{sec:results}
\begin{figure}[H]
\begin{subfigure}{0.55\textwidth}
   \includegraphics[page=1,width=\textwidth]{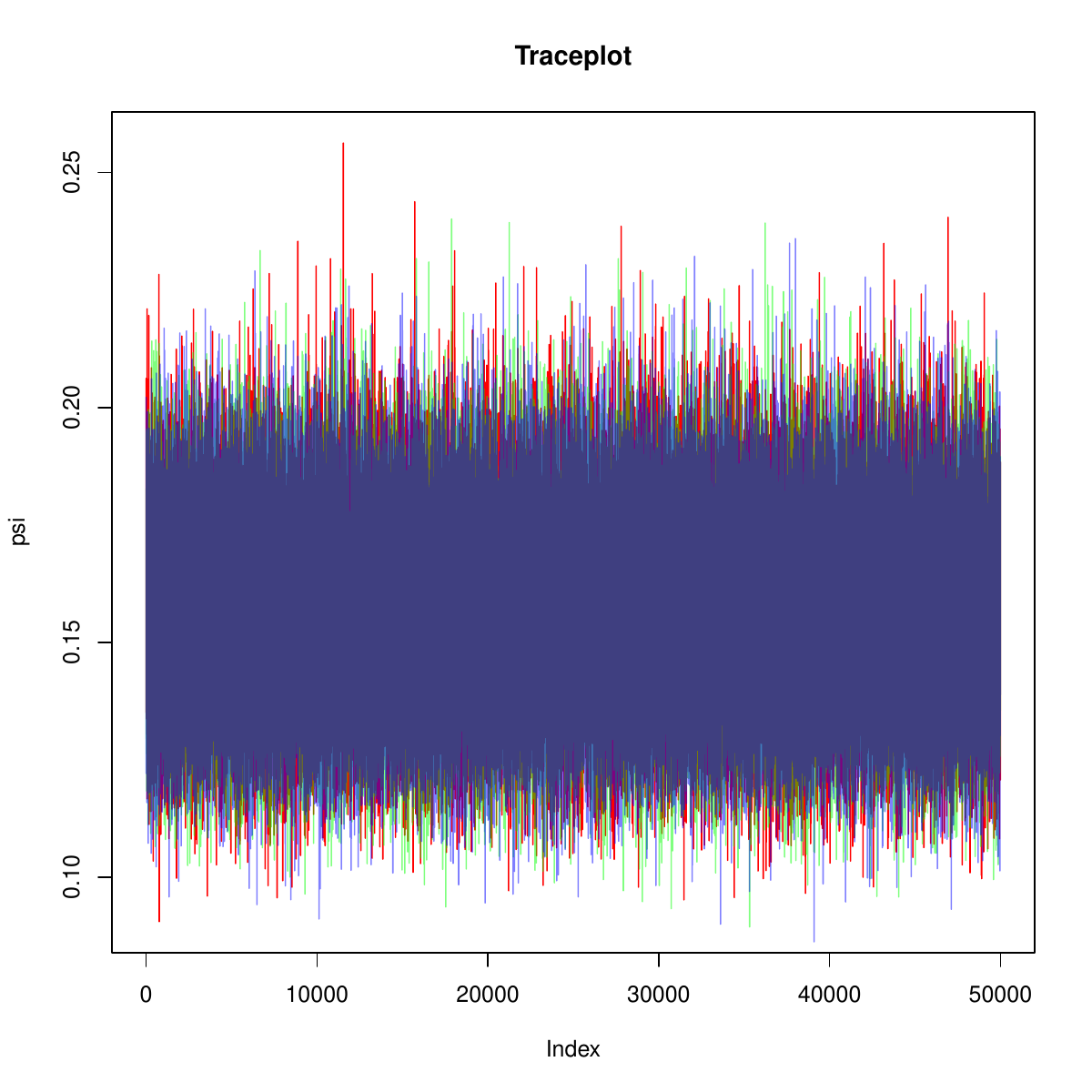}
   \caption{}
\end{subfigure}
\begin{subfigure}{0.55\textwidth}
     \includegraphics[page=2,width=\textwidth]{nu_1.5_S_NB_data.pdf}
     \caption{}
\end{subfigure}
\caption{(a) The traceplot $\psi$ based on three chains each of 50000 simulations for the model with tuning parameter $\nu=1.5$ using ISRO data (b) The density plot of $\psi$ obtained by kernel density estimation with Gaussian kernel}
\end{figure}

\begin{figure}[H]
\begin{subfigure}{0.55\textwidth}
   \includegraphics[page=3,width=\textwidth]{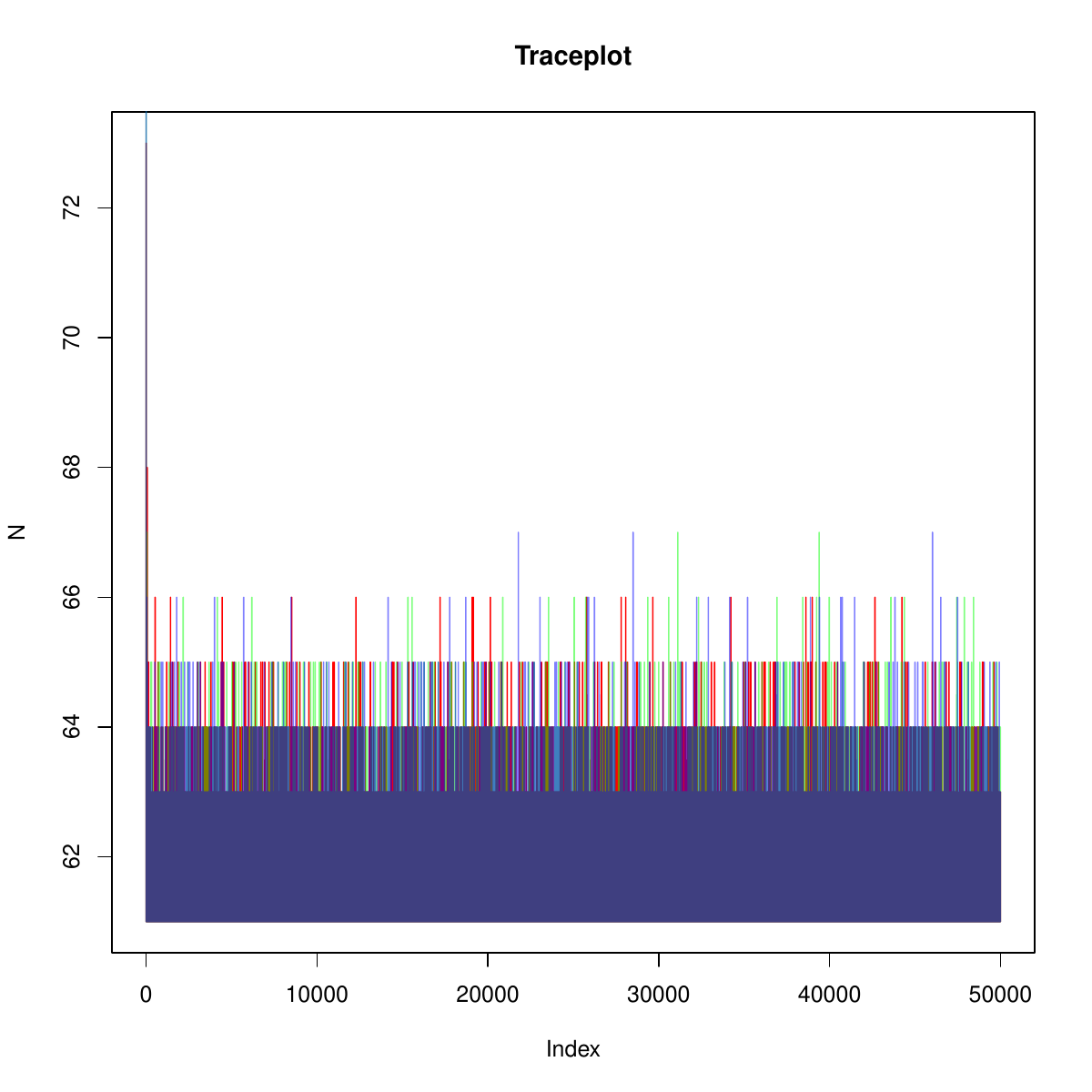}
   \caption{}
\end{subfigure}
\begin{subfigure}{0.55\textwidth}
     \includegraphics[page=4,width=\textwidth]{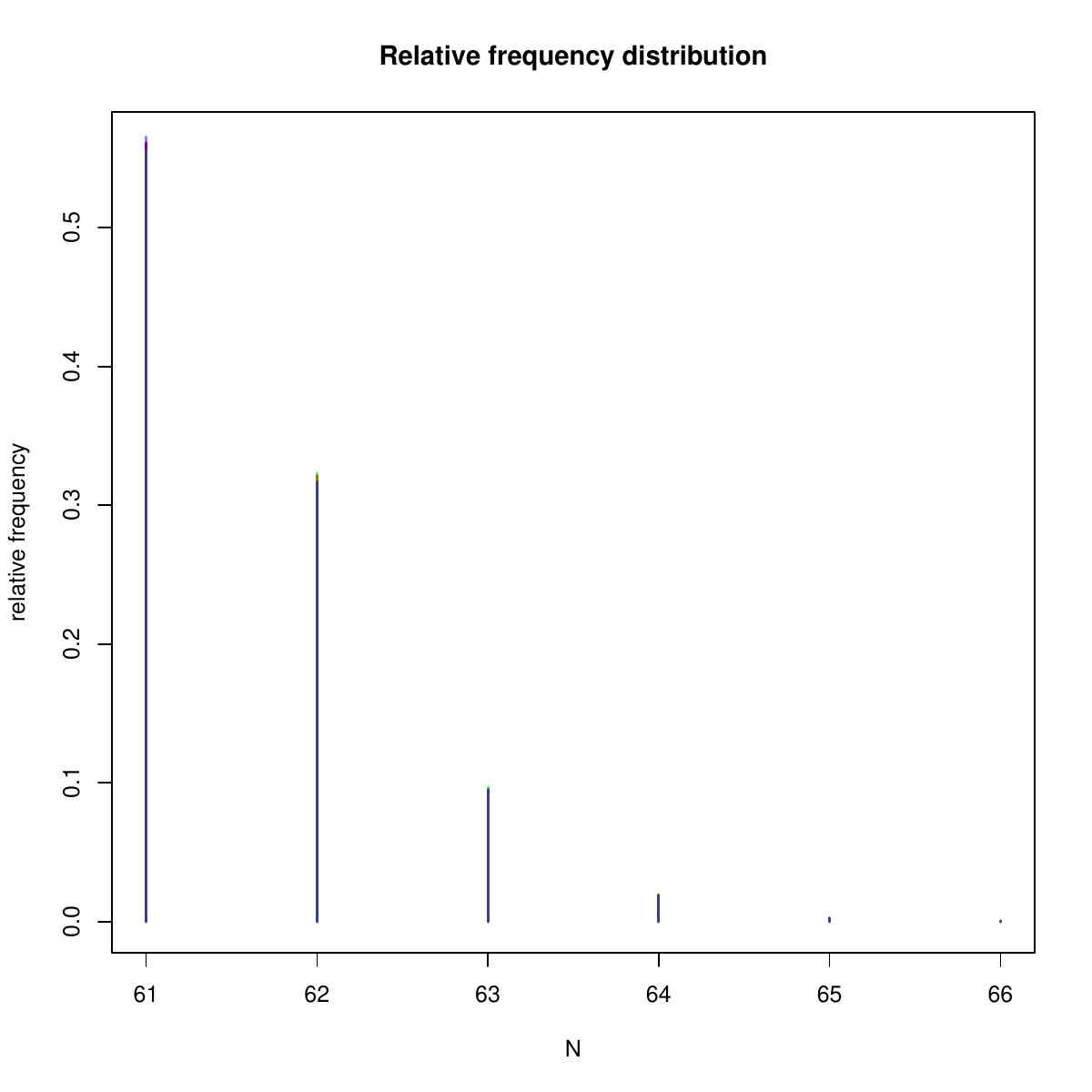}
     \caption{}
\end{subfigure}

\begin{subfigure}{0.55\textwidth}
   \includegraphics[page=15,width=\textwidth]{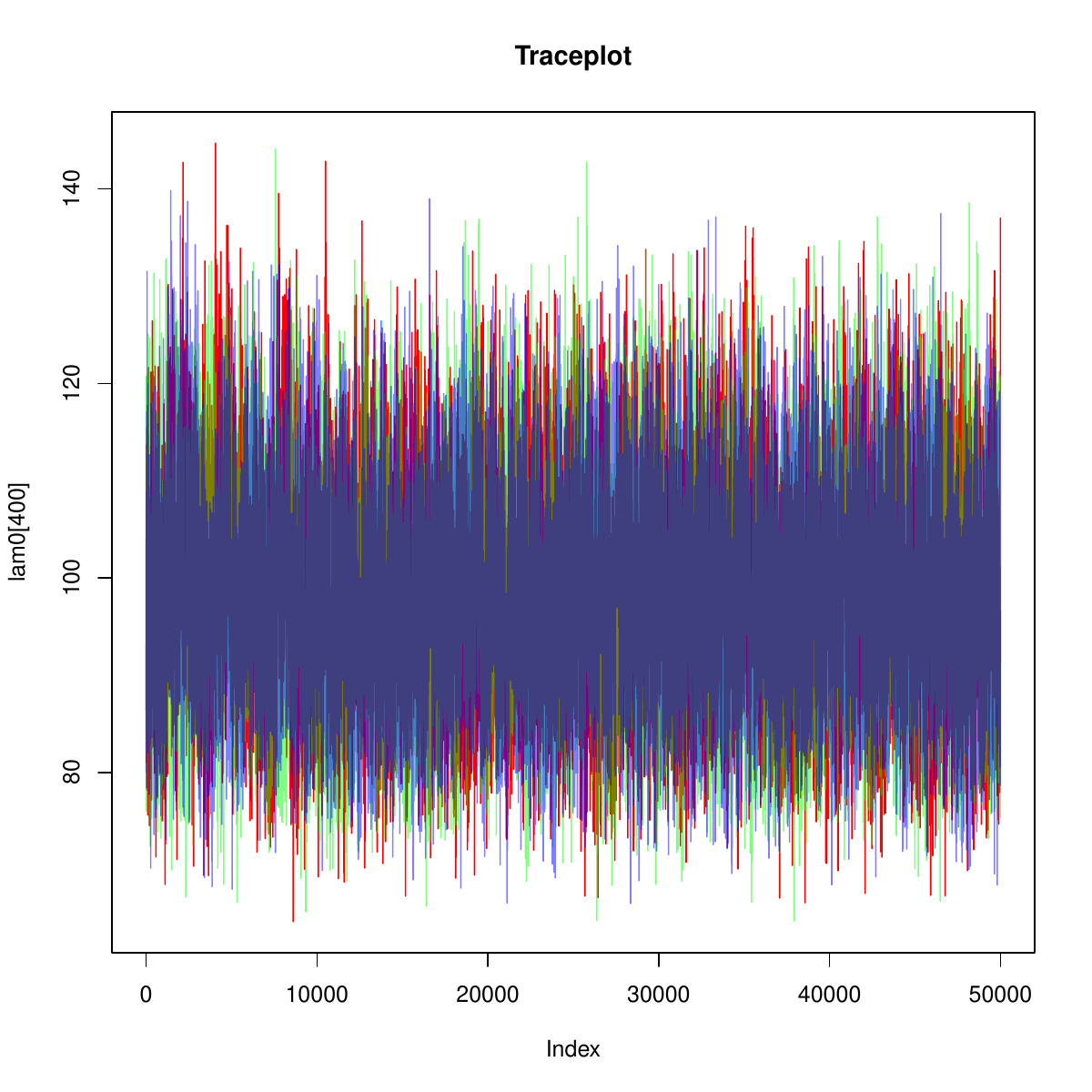}
   \caption{}
\end{subfigure}
\begin{subfigure}{0.55\textwidth}
     \includegraphics[page=16,width=\textwidth]{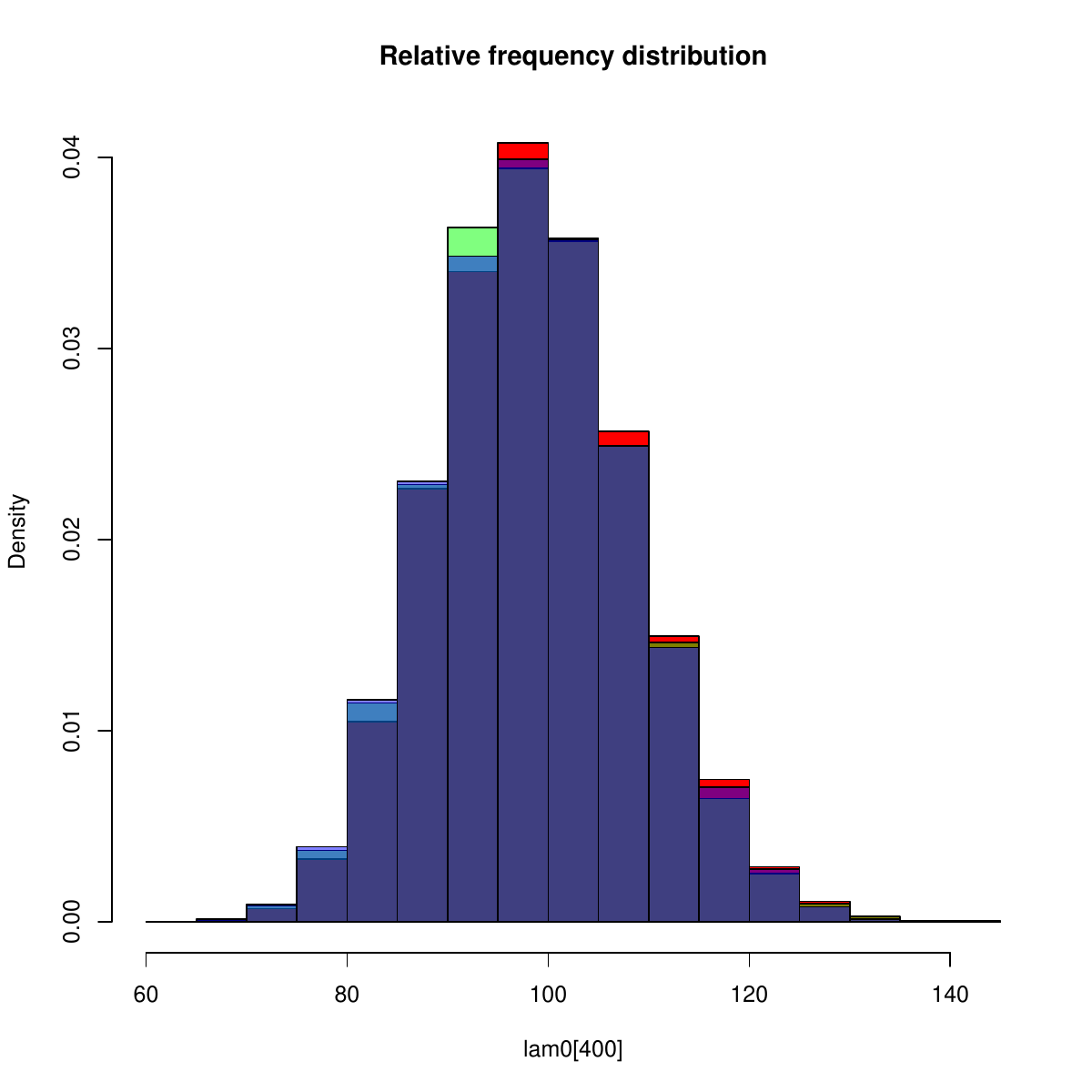}
     \caption{}
\end{subfigure}
\caption{(a) The traceplot of $N$ based on the three chains and 50000 repetitions for ISRO data (b) The sampling distribution of $N$ (c) The traceplot of $\lambda_{M}$ (d) The sampling distribution of $\lambda_{M}$. Here $M$ is set as 400}
\end{figure}

\begin{figure}[H]
\begin{subfigure}{0.55\textwidth}
   \includegraphics[page=27,width=\textwidth]{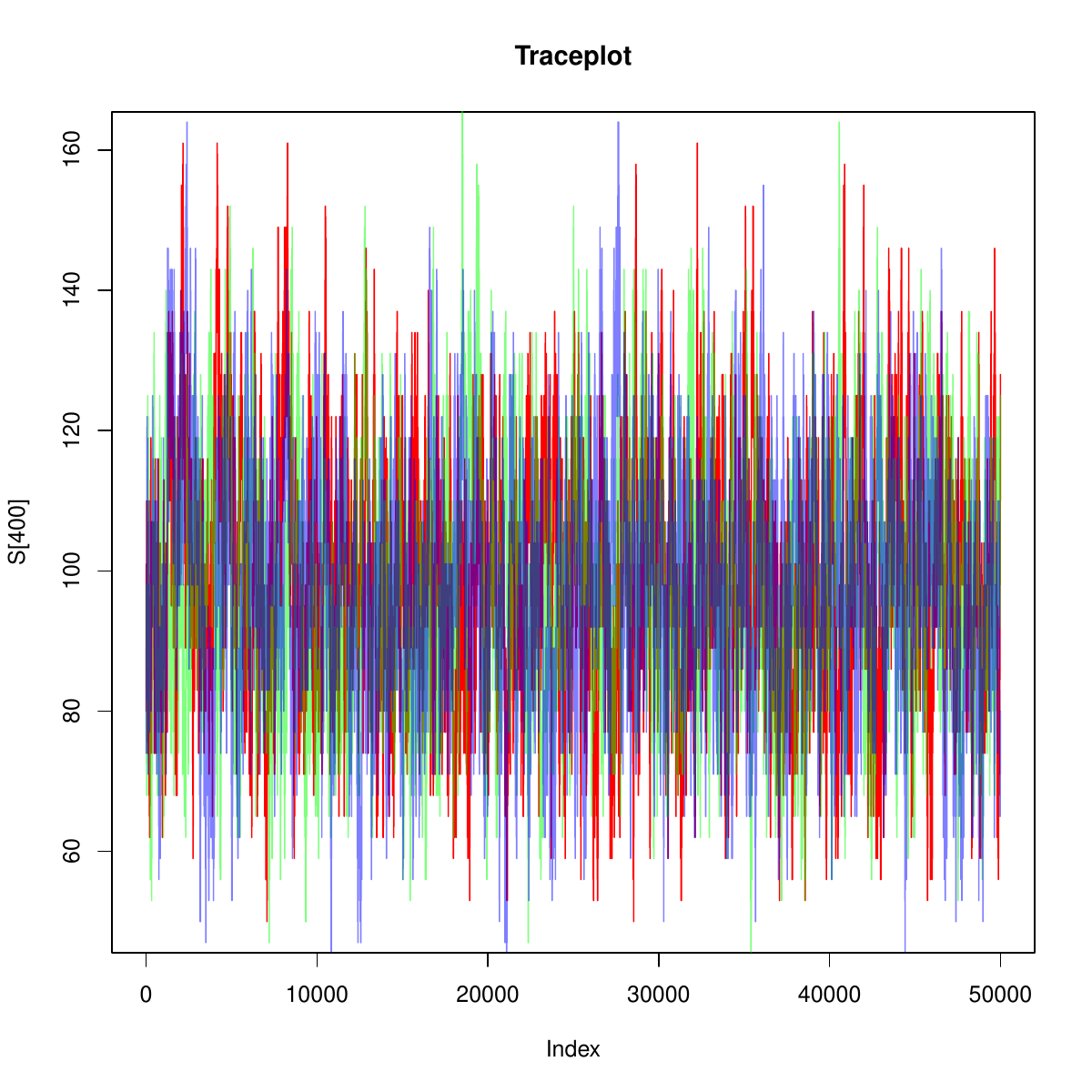}
   \caption{}
\end{subfigure}
\begin{subfigure}{0.55\textwidth}
     \includegraphics[page=28,width=\textwidth]{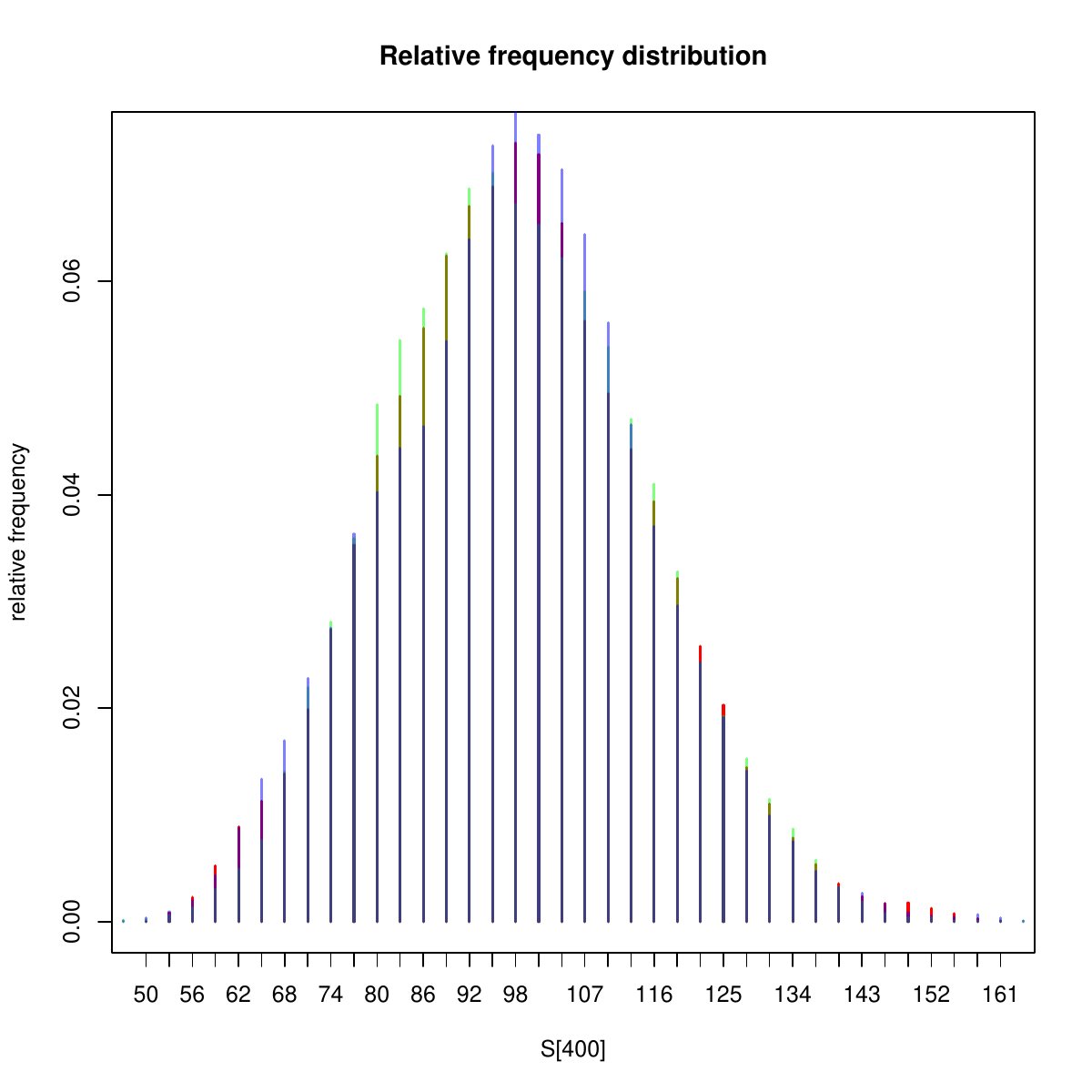}
     \caption{}
\end{subfigure}
\caption{(a) The traceplot $S_{M}$ based on three chains each of 50000 simulations for ISRO data (b) The sampling distribution of $S_{M}$. $M$ is set as 400.}
\end{figure}


The following table shows the descriptive measures for the model parameters:\vspace{2mm}
\\
\begin{table}[H]
    \centering
    \caption{Posterior summaries for each chain}
    \begin{tabular}{|p{1.09cm}|c|c|c|c|c|c|c|c|c|}
    \hline
        \multirow{2}{*}{para-} & \multicolumn{3}{c|}{mean} & \multicolumn{3}{c|}{sd} & \multicolumn{3}{c|}{cv}\\
         \cline{2-10}
         meters& ch1 & ch2 & ch3 & ch1 & ch2 & ch3 & ch1 & ch2 & ch3\\
         \hline
         $\psi$ & 0.1558 & 0.1557 & 0.1557 & 0.0182 & 0.0182 & 0.0181 & 11.71 & 11.66 & 11.60\\
         \hline
         $N$ & 61.5824 & 61.5915 & 61.5784 & 0.7678 & 0.7750 & 0.7703 & 1.25 & 1.26 & 1.25\\
         \hline
         $\lambda_{1}$ & 99.0490 & 99.1637 & 98.7765 & 10.1901 & 9.8482 & 9.8991 & 10.29 &  9.93 & 10.02\\
         \hline
         $\lambda_{2}$ & 99.0175 & 99.1637 & 98.7765 & 10.0106 & 9.8482 & 9.8991 & 10.11 &  9.93 & 10.02\\
         \hline
         $\lambda_{M-1}$ & 98.3540 & 99.0623 & 98.8662 & 9.8439 & 10.0914 & 9.8896 & 10.01 & 10.19 & 10.00\\
         \hline
         $\lambda_{M}$ & 98.7069 & 98.6535 & 98.6479 & 9.8872 & 9.8203 & 9.7837 & 10.02 &  9.95 & 9.92\\
         \hline
         $S_{1}$ & 99.3827 & 99.3363 & 98.3534 & 18.6644 & 16.8257 & 17.3005 & 18.78 & 16.94 & 17.59\\
         \hline
         $S_{2}$ & 99.2413 & 99.7911 & 99.6526 & 18.3529 & 16.6064 & 16.9403 & 18.49 & 16.64 & 17.00\\
         \hline
         $S_{M-1}$ & 97.1617 & 99.1214 & 98.2356 & 17.4022 & 18.3373 & 16.5193 & 17.91 & 18.50 & 16.82\\
         \hline
         $S_{M}$ & 98.0313 & 98.0200 & 98.0229 & 16.9898 & 16.4979 & 16.8775 & 17.33 & 16.83 & 17.22\\
         \hline
    \end{tabular}
    \label{tab:my_label5}
\end{table}

The estimates of the diagnostic measures to verify the convergence of the model parameters are tabulated below:\vspace{2mm}
\\
\begin{table}[H]
    \centering
    \caption{Diagnostic measures of convergence for different parameters}
    \begin{tabular}{|c|c|c|c|}
    \hline
         parameters & $\hat{R}$ & Upper C.I. & ESS\\
         \hline
         $\psi$ & 1 & 1 & 75253.50\\
         \hline
         $N$ & 1 & 1 & 70939.22\\
         \hline
         $\lambda_{M-2}$ & 1 & 1.01 & 1688.402\\
         \hline
         $\lambda_{M-1}$ & 1.01 & 1.03 & 1758.062\\
         \hline
         $\lambda_{M}$ & 1 & 1 & 1985.757\\
         \hline
         $S_{M-2}$ & 1.01 & 1.03 & 512.4497\\
         \hline
         $S_{M-1}$ & 1.02 & 1.08 & 515.3860\\
         \hline
         $S_{M}$ & 1 & 1 & 557.2808\\
         \hline
         
    \end{tabular}
    \label{tab:my_label6}
\end{table}
\noindent
Estimated eventual size of undetected bugs for the ISRO data = 138.3443. Estimated reliability of ISRO flight control software, based on the available data = 0.855 for $\eps=100$.r
From the traceplots, it is evident that the model worked well for the ISRO software data also. The table ~\ref{tab:my_label6} showing potential scale reduction factor ($\hat{R}$) for each parameter supports the convergence of the model. The coefficient of variation for each parameter is small enough indicating the high accuracy of the model parameters. Using posterior mean, we estimate the total number of bugs present in the software as 62 with $95\%$ credible interval (61, 63) which is quite precise. Since on the basis of our estimation we might have one more bug in the software, we can consider the minimum value of $\eps$ as 100. However, for values above hundred the reliability estimates are shown in Table~\ref{tab:my_label7}. 

\begin{table}[H]
    \centering
    \caption{Reliability estimates for different threshold values ($\eps$)}
    \begin{tabular}{|p{3cm}|p{4cm}|}
        \hline
        $\eps$ & Reliability\\
        \hline
         100 & 0.8539792\\
         \hline
         120 & 0.8828314\\
         \hline
         140 & 0.9004306\\
         \hline
         160 & 0.9329227\\
         \hline
         180 & 0.9635215\\
         \hline
         200 & 0.9763476\\
         \hline
    \end{tabular}
    \label{tab:my_label7}
\end{table}
\section{Discussion}\label{sec:discussion}
Here we develop a model depending on a newly introduced concept of size of a bug. It is clear that a bug in a rare path may be undetected by the preassigned test cases. So, it may be assumed that a very few test cases can pass through the particular bug. Therefore, the eventual size of such bugs should be very small. If the eventual size of the remaining bugs is small enough, we can infer that the software is highly reliable. So, we target to build a model to find out the eventual size of the remaining bugs and hence the reliability of the software in this article.

\par For developing our model, we adopted a particular procedure of testing used for critical software like ISRO data. The softwares have been tested through several phases with some prespecified inputs (test-cases). The data has been collected for several space missions. For modelling we assume that a bug can be detected in any phase of any mission. However a detected bug in a phase is debugged before the testing for next phase starts. So, while building the model, we consider the number of test-cases used in each phase and the bugs detected in the same phase. We use Bayesian method to estimate the model parameters. We validate the model through simulation study. A tuning parameter is used to determine the probability of non-detection. Finally, we observe the posterior means, standard deviation and coefficient of variation of the parameters for different values of tuning parameters to decide the model. We also find the model as highly reliable for the chosen tuning parameter values of $1.0$, $1.25$ and $1.50$.

\par Subsequently, we apply our model to the ISRO data and we found a satisfactory result by this model. From the ISRO data, it is to be noted that, 61 bugs were detected in 35 missions each with 8 phases. Whereas, we estimate the total number of bugs in the software approximately as 62 and the reliability as 0.85. 
 
\subsection{Conclusions}\label{sec:conclusions}
The concept of eventual size of a bug in a very recent one (\cite{dey2022estimating}). In our model we used this very new concept to estimate reliability of some critical space application software. The model along with the eventual size concept may be generalized to other similar fields like hydro carbon exploration \cite{Chakraborty1994Optimum}. Our proposed model could estimate the reliability of ISRO software with a high degree of confidence, since the $95\%$ credible interval turns out to be (61,63), whereas the actual observed number of bugs being 61. 

\par In Bayesian analysis, selection of priors play an important role. We have used non-informative priors for the estimation so that the prior knowledge on the parameters are restricted. For some parameters, different priors have been used to check model robustness. Though it is desirable to conduct sensitivity analysis before one applies the model.\vspace{6mm}
{\bf Conflicts of interest}: It is hereby declared that the authors have no conflicts of interest.\\
{\bf Availability of data and material}: The data used in the article, R codes for simulation study and application using data are provided in GitHub. \href{https://github.com/pg-stat/Software-reliability}{https://github.com/pg-stat/Software-reliability} \\


\bibliographystyle{plainnat}
\section{Appendix}
\href{https://drive.google.com/file/d/1jtPeyZKF1F7Gy5oT8REZ9Jv75NBnosxz/view?usp=drive_link}{Appendix I: simulation study}\vspace{2mm}
\\
\href{https://drive.google.com/file/d/1v6mi4_XMvOiUoAJy6nBP4ui9DpX_4Wb9/view?usp=drive_link}{Appendix II: ISRO data}

\bibliography{bibliography}

\end{document}